# Magneto-caloric effect in the pseudo-binary intermetallic YPrFe$_{17}$ compound

Pablo Álvarez[a], Pedro Gorria[a,*], José L. Sánchez Llamazares[b], María J. Pérez[a], Victorino Franco[c], Marian Reiffers[d], Jozef Kováč[d], Inés Puente-Orench[e], Jesús A. Blanco[a].

[a]*Departamento de Física, Universidad de Oviedo, Calvo Sotelo, s/n, 33007 Oviedo, Spain*

[b]*División de Materiales Avanzados, Instituto Potosino de Investigación Científica y Tecnológica, Camino a la presa San José 2055, CP 78216, San Luis Potosí, Mexico.*

[c]*Departamento de Física de la Materia Condensada, ICMSE-CSIC, Universidad de Sevilla, P.O. Box 1065, 41080 Sevilla, Spain*

[d]*Institute of Experimental Physics, Watsonova 47, SK-04001 Košice, Slovakia.*

[e]*Institute Laue Langevin, 6 rue Jules Horowitz, 38042 Grenoble, France*

*E-mail: pgorria@uniovi.es

**Abstract.** We have synthesized the intermetallic YPrFe$_{17}$ compound by arc-melting. X-ray and neutron powder diffraction show that the crystal structure is rhombohedral with space group $R\bar{3}m$ (Th$_2$Zn$_{17}$-type). The investigated compound exhibits a broad magnetic entropy change $\Delta S_M(T)$ associated with the ferro-to-paramagnetic phase transition ($T_C \approx 290$ K). The isothermal $|\Delta S_M|$ ($\approx 2.3$ J kg$^{-1}$ K$^{-1}$) and the relative cooling power ($\approx 100$ J kg$^{-1}$) have been calculated for applied magnetic field changes up to 1.5 T. A single master curve for $\Delta S_M$ under different values of the magnetic field change can be obtained by a rescaling of the temperature axis. The results are compared and discussed in terms of the magneto-caloric effect in the isostructural R$_2$Fe$_{17}$ (R = Y, Pr and Nd) binary intermetallic alloys.

*Keywords:*   A. Magnetically ordered materials; C. Crystal structure and symmetry; D. Magneto-caloric effect; E. Magnetic measurements

## 1. Introduction

The magneto-caloric effect (MCE) is nowadays a subject of considerable current research interest [1-3] motivated by the enhanced performance (efficiency, mechanical vibration, size, etc) and reduced environmental impact of refrigeration systems based on this effect compared with those of the existing vapour-compression gas technology [4]. In this way, diverse types of compounds have been investigated until now with the aim of scrutinizing both the intensity and the temperature range for the MCE [3, 5-7]. A magnetic material must exhibit a large magnetization change, $\Delta M$, around its magnetic phase transition in order to display a large MCE response. From the point of view of the implementation of these materials in magnetic refrigeration systems, another important parameter is the relative cooling power (*RCP*) [8,9], which gives a figure of merit of how much heat could be transferred between the hot and cold reservoirs by the magnetic refrigerant in an ideal thermodynamic cycle. For materials with second-order phase transition the peak value of the magnetic entropy change $\left|\Delta S_M^{peak}\right|$ is smaller than that observed in materials with first-order magnetic phase transitions [1,2]. Nevertheless the lack of magnetic field hysteresis and the larger operation temperature range in materials displaying a second-order magnetic phase transition give commonly rise to higher *RCP* values [9-11]. Therefore, a compromise between the magnitude of the magnetic entropy change and the *RCP* is mandatory for employing a magnetic material in magnetic refrigeration applications. Mostly of the current prototypes for room temperature magnetic refrigeration employ rare-earth-based materials (with the rare-earth being mainly Gd) [2,12]. However, some Fe-rich $R_2Fe_{17}$ (R = Rare Earth) compounds have shown values of *RCP* comparable with those of Gd-based magnetic materials together with a lower cost of the main component (Fe), easy fabrication procedures and absence of disadvantageous hysteresis effects [13,14]. Within the whole $R_2Fe_{17}$ series the alloys with R = Y, Pr or Nd have the largest magnetic moment per formula unit, and therefore the higher $\left|\Delta S_M^{peak}\right|$ value [15], due to the collinear ferromagnetic order, and also magnetic ordering temperatures, $T_C$, around 300 K. These facts make them suitable for their use in magnetic refrigeration as active magnetic regenerative systems operating around room temperature [16]. Moreover, the value of $T_C$ in this 2:17 type of compounds can be tuned by mixing two rare earth elements (i.e. in the form $R_{2-x}R'_xFe_{17}$) [17], or by partial substitution of Fe other 3d-atom [18]. The crystal structure of the binary intermetallic $R_2Fe_{17}$ compounds can be either of $Th_2Zn_{17}$-type (rhombohedral $R\overline{3}m$ space group) for light rare-earths, or $Th_2Ni_{17}$-type (hexagonal $P/6_3mmc$ space group) for heavy rare-earths [19,20]. In the case of R = Y, Gd and Tb or in pseudo-binary intermetallic $R_{2-x}R'_xFe_{17}$ alloys with a mixture of two different R atoms, both crystal structures can coexist, with the rare-earths sharing the crystallographic sites [21-24].

In the present work we have studied the crystal structure of a new pseudo-binary intermetallic $YPrFe_{17}$ alloy by means of neutron and x-ray powder diffraction, together with its magnetic properties and the magneto-caloric effect up to a maximum applied magnetic field change of $\mu_0\Delta H$ = 1.5 T. The experimental results are compared with those measured in binary $Y_2Fe_{17}$, $Pr_2Fe_{17}$ and $Nd_2Fe_{17}$.

## 2. Experimental details and data analysis

As-cast ingots with $YPrFe_{17}$ nominal composition were prepared from 99.99% pure elements (relative to rare earth content in the case of Pr) by standard arc-melting technique under a controlled Ar atmosphere. The polycrystalline as-cast pellets were sealed under vacuum in quartz ampoules and further annealed during one week at 1263 K. After finishing the heat treatment the samples were quenched directly in water. Crystal structure was determined at room temperature (*T* = 290 K) by both x-ray (XRD) and neutron powder diffraction (ND). XRD studies were performed in a high-resolution x-ray powder diffractometer (Seifert model XRD3000) operating in Bragg-Bentano geometry. The scans in *2θ* were performed between 2 and 160º with 0.02º steps and counting times of 20 s per point using *Cu K*$_α$ radiation ($\lambda$ = 1.5418 Å). The ND pattern was collected on the high-intensity D1B two-axis powder diffractometer at the ILL (Grenoble) with a neutron wavelength of $\lambda$ = 2.52 Å, 2 hours of acquisition time and an angular range of 80º in *2θ* (in steps of 0.2º). The full-profile analysis of the diffraction patterns was carried out with the FullProf suite package [25], and no peak broadening due to small crystal and/or microstrain effects [26] were detected.

The low-magnetic field magnetization as a function of temperature *M(T)* curves were recorded in a Faraday susceptometer under a heating rate of 2 K/min. Isothermal magnetization curves, *M(H)*, were measured with a Lakeshore model 7407 VSM vibrating sample magnetometer in the temperature range between 90 and 450 K with a maximum applied magnetic field of 1.5 T, and in a Quantum Design MPMS-5T magnetometer in the temperature range 90 – 390 K with applied magnetic fields up to 5 T. At each temperature the magnetization was measured for a large number selected values of the applied magnetic field ($\approx$ 150 for the VSM measurements and 50 for the MPMS) with the aim of gaining accuracy in the estimation of the isothermal magnetic entropy change, $|\Delta S_M|$. The value of $|\Delta S_M|$ at each temperature *T* due to a change of the applied magnetic field from *H* = 0 to *H* = $H_{max}$ were calculated using the Maxwell relation [8]:

$$\Delta S_M(T,H) = S_M(T,H) - S_M(T,0) = \int_0^H \left(\frac{\partial M(T',H')}{\partial T'}\right)_{T'=T} dH' \qquad (1)$$

After applying this procedure to the whole set of $M(H)$ curves the value of $|\Delta S_M|$ for a given applied magnetic field change and at a selected temperature is obtained by numerical approximation of eq. 1, where the partial derivative is replaced by finite differences and then the integral is calculated by means of numerical methods [6,9]. In addition, the relative cooling power ($RCP$) has been calculated using three different criteria (see reference [9] for details): $RCP\text{-}1(H)$ = $|\Delta S_M^{max}|(H) \times \delta T_{FWHM}(H)$, where $\delta T_{FWHM}$ is the full width at half maximum of $|\Delta S_M|(T)$ curve; $RCP\text{-}2$ is the area below $|\Delta S_M|(T)$ curve in the temperature range between $T-\delta T_{FWHM}$ and $T+\delta T_{FWHM}$; and $RCP\text{-}3$ is the maximum value of the product $|\Delta S_M|\times \Delta T$ below the $|\Delta S_M(T)|$ curve.

## 3. Results and discussion

In Fig. 1 the room temperature x-ray (upper panel) and neutron (bottom panel) powder diffraction patterns of the YPrFe$_{17}$ sample are shown.

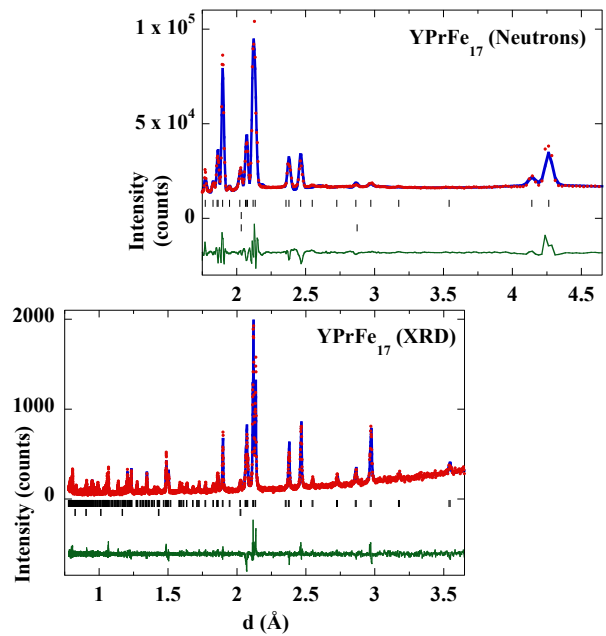

**Fig. 1.** Observed (dots) and calculated (solid line) powder diffraction patterns for YPrFe$_{17}$ alloy collected at $T = 300$ K. Positions of the Bragg reflections are represented by vertical bars; the first row corresponds to the rhombohedral Th$_2$Zn$_{17}$–type phase while the second one is associated with an α-Fe impurity (< 3%). The observed–calculated difference is depicted at the bottom of each figure.

Both XRD and neutron diffraction patterns have been refined by using the Rietveld method in multi-pattern mode (see [25] for further technical details). The observed intensity peaks can be indexed as the Bragg reflections corresponding to a rhombohedral Th$_2$Zn$_{17}$-type crystal structure with $R\overline{3}m$ space group (#164), and if the hexagonal setting is chosen the lattice parameters are: $a$ = 8.540 (1) Å and $c$ = 12.419 (1) Å. No traces of the hexagonal Th$_2$Ni$_{17}$-type crystal structure have been found, as confirmed by neutron diffraction from which information of the whole sample is attained, in contrast with XRD. Whereas in other 2:17 pseudo-binary alloys of this family a disordered rhombohedral structure has been proposed in order to explain the strong intensity reduction observed for many reflexions [27,28], in the present case there is no such a reduction, even though when compared with the diffraction patterns of the binary Y$_2$Fe$_{17}$, Pr$_2$Fe$_{17}$ or Nd$_2$Fe$_{17}$ alloys [13,29]. From the fit of the diffraction patterns it is evident that the 6c site corresponding to the R atoms is equally shared between Y and Pr with the same atomic coordinates. The values for the main crystallographic parameters are given in Table I. The values for the cell parameters are, as it could be expected, between those of the $Y_2Fe_{17}$ and $Pr_2Fe_{17}$ ($a$ = 8.46 Å, $c$ = 12.39 Å and $a$ = 8.585 Å; $c$ = 12.464 Å, respectively) [13,20,30]. The ratio c/a ~ 1.54 is in good agreement with those reported for the rhombohedral crystal structure in these 2:17-type alloys [13,29,30].

In order to estimate the Curie temperature of the sample, the magnetization vs. temperature, $M(T)$, curve under a low applied magnetic field $\mu_0H$ = 5 mT (not shown) has been measured. Thereafter, the value of $T_C$ has been taken as the minimum of the $dM/dT$ vs. $T$ curve, which is a commonly adopted criterion [13,24,29]. In this way, $T_C$ = 291 ± 5 K for YPrFe$_{17}$, which is in between those reported values for $Pr_2Fe_{17}$ ($T_C$ = 286 ± 2 K [13]), and for $Y_2Fe_{17}$ ($T_C$ = 301 ± 4 K [15]). Therefore, it seems that by mixing Y and Pr the Curie temperature of the resulting pseudo-binary alloy can be tuned between those values of the pure binary compounds.

Fig. 2 shows a 3D surface plot representing simultaneously the temperature and magnetic field dependences of the magnetization, $M(H,T)$ for YPrFe$_{17}$ alloy.

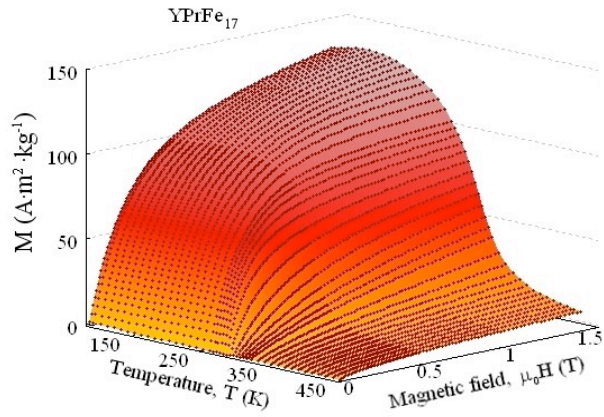

**Fig. 2.** 3D surface corresponding to the temperature and applied magnetic field dependences for the magnetization of the YPrFe$_{17}$ compound.

The isothermal magnetic entropy change, $|\Delta S_M|$ has been calculated from the set of isothermal magnetization vs. applied magnetic field, $M(H)$, curves depicted in Fig. 2 and following the procedure explained in the previous section. In Fig. 3 the temperature dependence of $|\Delta S_M|$ for the maximum applied magnetic field change, from 0 to $\mu_0H_{max}$ = 1.5 T is shown. In addition, data for $Y_2Fe_{17}$, $Pr_2Fe_{17}$ and $Nd_2Fe_{17}$ binary intermetallic compounds with the same rhombohedral $Th_2Zn_{17}$-type crystal structure are also shown for comparison. The maximum value for the magnetic entropy change, $\left|\Delta S_M^{peak}\right|$, for the YPrFe$_{17}$ compound is 2.3 J kg$^{-1}$ K$^{-1}$, which is just in between those values for the binary $Pr_2Fe_{17}$ (2.6 J K$^{-1}$ kg$^{-1}$) and $Y_2Fe_{17}$ (1.9 J K$^{-1}$ kg$^{-1}$) alloys. The latter can be understood taking into account that $\left|\Delta S_M^{peak}\right|$ is roughly proportional to the magnetization change of the alloy across the second order ferro- to paramagnetic phase transition. In the case of $Pr_2Fe_{17}$, the magnetic moments of Pr and Fe sublattices are parallel to each other, hence, the contribution of Pr atoms (≈ 3 $\mu_B$/Pr atom [31]) to the net magnetization of the alloy is additive, while in the case of $Y_2Fe_{17}$, Yttrium atoms do not carry any magnetic moment, and the net magnetization of the alloys comes exclusively from the Fe sublattice. Assuming that: (i) the Fe atoms possess the same values for the magnetic moment in $Pr_2Fe_{17}$, YPrFe$_{17}$ and $Y_2Fe_{17}$ alloys; (ii) the $M(T)$ curves show a very similar trend for the three alloys; and (iii) the $Pr^{+3}$ ions in YPrFe$_{17}$ have their magnetic moments parallel to those of Fe atoms, we could expect that the substitution of half of the $Pr^{+3}$ ions by Y ones should give rise to a decrease in the magnetic entropy change, respect to that of $Pr_2Fe_{17}$, down to an approximate value given by: $\left|\Delta S_M\right|_{YPrFe_{17}} \approx \frac{1}{2}\left(\left|\Delta S_M\right|_{Pr_2Fe_{17}} + \left|\Delta S_M\right|_{Y_2Fe_{17}}\right)$.

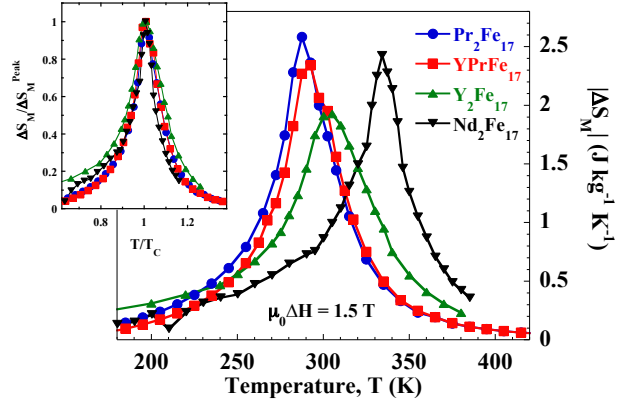

**Fig. 3.** Temperature dependence of the magnetic entropy change in the pseudo-binary YPrFe$_{17}$ compound around the |$\Delta S_M$| peak for an applied magnetic field change $\mu_0\Delta H$ = 1.5 T. Data for binary R$_2$Fe$_{17}$ (R = Y, Pr and Nd) are also shown for comparison. The lines connecting the calculated points are guides for the eyes. Inset: Normalized $\Delta S_M/\Delta S_M^{Peak}$ vs. $T/T_C$ for the four intermetallic alloys.

We summarize in table 2 the values for $T_C$, |$\Delta S_M$| (peak value for $\mu_0\Delta H$ = 1.5 T) together with the *RCP* estimated by using the three criteria previously defined. It is worth noting that although Y$_2$Fe$_{17}$ exhibits the lowest |$\Delta S_M$| the *RCP* -1 is the highest due to a broader |$\Delta S_M$|(T) peak as it can be observed in the inset of Fig. 3, where |$\Delta S_M$| vs. the reduced temperature $T/T_C$ is plotted.

In Fig. 4 the magnetic field dependence of the *RCP* for YPrFe$_{17}$ is depicted. From a linear fit of the *RCP* (*H*) curves for $\mu_0 H$ > 1 T, we have extrapolated the values for an applied magnetic field change of 2 T in order to compare with available date for Gd (see table 2), which is the archetypical magento-caloric material with second order magnetic phase transitions.

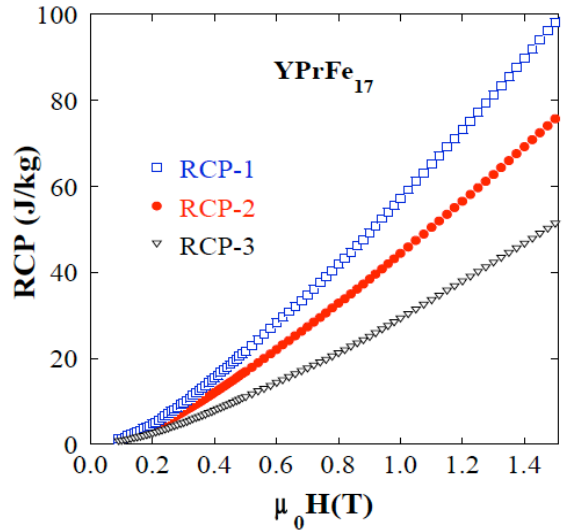

Figure 4. Magnetic field dependence of the Relative Cooling Power (*RCP*). See text for details.

The pseudo-binary YPrFe$_{17}$ alloy exhibits *RCP* values comparable to those of the R$_2$Fe$_{17}$ compounds with R = Y, Pr or Nd), and both *RCP* -1 and *RCP* -2 are ca. 75 % of those for pure Gd [32].

Moreover, it has been proposed that |$\Delta S_M$|(T) curves for different magnetic field changes can collapse into a single master curve, after an appropriate normalization, in ferromagnetic materials exhibiting a second order magnetic phase transition [33]. The master curve is obtained as follow [34]: firstly, the |$\Delta S_M$|(T) curves are normalized to its maximum value |$\Delta S_M^{max}$| for each value of the applied magnetic field change. Secondly, the temperature axis is rescaled

using two different reference temperatures:

$$\theta = -(T - T_C)/(T_{r1} - T_C) \qquad T < T_C \qquad (2)$$

$$\theta = (T - T_C)/(T_{r2} - T_C) \qquad T > T_C \qquad (3)$$

where $T_{r1}$ and $T_{r2}$ are the temperatures at which $|\Delta S_M| = a \times |\Delta S_M^{Peak}|$, with $0 \leq a \leq 1$. In our case we have chosen $a = 0.5$ (this value makes $T_{r1}$ and $T_{r2}$ coincident with those temperatures at which $|\Delta S_M| = |\Delta S_M^{Peak}|/2$ [34]). The master curves for the YPrFe$_{17}$ compounds are shown in Fig. 5, and it can be observed how the $\Delta S_M/\Delta S_M^{Peak}$ vs. $\theta$ curves for different magnetic field values almost collapse into a unique one for each compound.

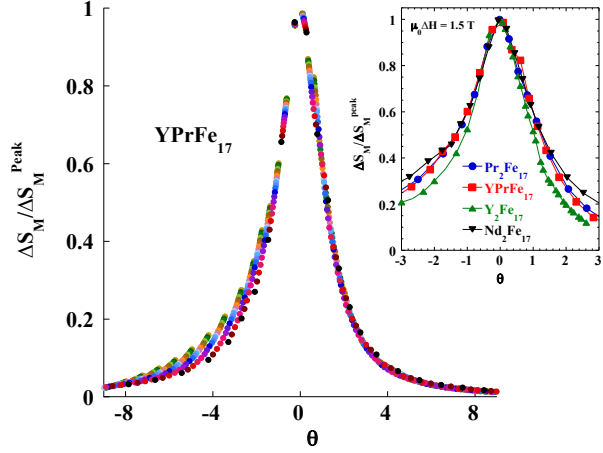

**Fig. 5.** Normalized $|\Delta S_M|$ vs. reduced temperature for the pseudo-binary YPrFe$_{17}$ compound. The inset shows the comparison with the curves for the binary Y$_2$Fe$_{17}$, Pr$_2$Fe$_{17}$ and Nd$_2$Fe$_{17}$ for an applied magnetic field $\mu_0 H = 1.5$ T.

Moreover, it can be shown that if the $\Delta S_M/\Delta S_M^{Peak}$ vs. $\theta$ curve for YPrFe$_{17}$ compound, corresponding to $\mu_0 \Delta H = 1.5$ T, is compared with those for the binary R$_2$Fe$_{17}$ alloys (R = Y, Pr, Nd, see inset in Fig. 5), the curves almost overlap in the range $-1 \leq \theta \leq 1$ (for temperatures between $T_{r1}$ and $T_{r2}$), thus suggesting that in these compounds the temperature dependence of the magnetic entropy change exhibits a similar trend [35]. From a practical point of view, the building of such a master curve can be considered as an useful tool for extrapolating the $|\Delta S_M|(H,T)$ curves for temperature and/or applied magnetic field values different from those available in the laboratory [36]. On the other hand, it also could help in detecting and studying, from a more fundamental viewpoint, different co-existing magnetic phenomena [37].

**Summary and conclusions**

The magnetic properties and the magneto-caloric effect in the pseudo-binary YPrFe$_{17}$ alloy have been studied. Room temperature x-ray and neutron powder diffraction confirm that the compound crystallizes into the ordered Th$_2$Zn$_{17}$-type rhombohedral crystal structure. The magnetic entropy change has been obtained by the isothermal magnetic measurements, showing that the introduction of the non-magnetic Y atoms leads to a shift of the temperature where the maximum of $|\Delta S_M|$ is obtained with a small reduction of the peak value. The calculated values for the *RCP* in YPrFe$_{17}$ can reach 75 % of the pure Gd, Therefore, we could expect that an adequate mixture of Pr or Nd with Y in R$_2$Fe$_{17}$ compounds allows us tuning the Curie temperature around room temperature (between 285 and 340 K) with almost similar values for the *RCP*. Therefore, (YPrNd)$_2$Fe$_{17}$ compounds are potential candidates for its use in magnetic refrigeration. Finally, the magnetic entropy change for several magnetic fields can be represented using a master curve representation for all these alloys.

**Acknowledgments**

We thank Spanish MICINN and FEDER programme for financial support through the research project MAT2008-06542-C04-03 and the Slovak grant agency VEGA 2/0007/09. P.A. is grateful to FICyT for Ph.D. contract. The Slovak Research and Development Agency (contract No. VVCE-0058-07), the CLTP as the Centre of Excellence SAS and P.J. Šafárik University, the CEX Nanofluid as the Centre of Excelence SAS, the 7.FP EU–MICROKELVIN and the SCT's at the University of Oviedo (XRD measurements) are also acknowledged. We also thank ILL and Spanish CRG-D1B for allocating neutron beam time.

Table 1. Table I. Crystallographic parameters, cell volume and atomic coordinates of the studied $R_2Fe_{17}$ ($R\bar{3}m$) compounds obtained from the both NPD and XRD patterns in multi-pattern fit.

| | |
|---|---|
| a (Å) | 8.540 (1) |
| c (Å) | 12.419 (1) |
| c/a | 1.454 |
| V (Å$^3$) | 784.3 (2) |
| Pr/Y (6c) | |
| z | 0.348 (3) |
| Fe1 (6c) | |
| z | 0.092 (1) |
| Fe3 (18f) | |
| x | 0.293 (1) |
| Fe4 (18h) | |
| x | 0.169 (2) |
| z | 0.489 (1) |
| $R_B$ | 5.7 |
| $\chi^2$ (%) | 1.5 |

Table 2. Curie temperature, $T_C$, magnetic entropy change, $|\Delta S_M|$ and relative cooling power, $RCP$, obtained from the three methods (see text). Extrapolated values of $RCP$ for a magnetic field change $\mu_0\Delta H = 2$ T are compared with those for Gd taken from Ref. [32].

| **Alloy** | YPrFe$_{17}$ | Pr$_2$Fe$_{17}$ | Y$_2$Fe$_{17}$ | Nd$_2$Fe$_{17}$ | Gd |
|---|---|---|---|---|---|
| $T_C$ (K) | 290(5) | 286(2) | 303(4) | 339(2) | 291(2) |
| $|\Delta S_M|$ (1.5 T) (J·kg$^{-1}$·K$^{-1}$) | 2.3 | 2.6 | 1.9 | 2.5 | --- |
| $T_{r1}$ (K) | 269 | 268 | 278 | 314 | --- |
| $T_{r2}$ (K) | 312 | 308 | 334 | 349 | --- |
| **RCP-1** (1.5 T) (J kg$^{-1}$) | 98 | 101 | 112 | 85 | --- |
| **RCP-2** (1.5 T) (J kg$^{-1}$) | 75 | 78 | 86 | 64 | --- |
| **RCP-3** (1.5 T) (J kg$^{-1}$) | 51 | 55 | 56 | 57 | --- |
| **RCP-1** (2 T) (J kg$^{-1}$) | 145 | 156 | 155 | 133 | 200 |
| **RCP-2** (2 T) (J kg$^{-1}$) | 111 | 116 | 121 | 98 | 147 |
| **RCP-3** (2 T) (J kg$^{-1}$) | 77 | 81 | 82 | 81 | 135 |